\documentclass[prb,aps,twocolumn,showpacs,floats]{revtex4}
\usepackage{epsfig,bm,epsf,graphics}
\usepackage{amsmath}
\begin{document}
\title{Temperature Dependence of the Spin Resistivity  in Ferromagnetic
Thin Films}
\author{K. Akabli and  H. T. Diep\footnote{ Corresponding author, E-mail:
diep@u-cergy.fr}}
\affiliation{Laboratoire de Physique Th\'eorique
et Mod\'elisation,
CNRS-Universit\'e de Cergy-Pontoise, UMR 8089\\
2, Avenue Adolphe Chauvin, 95302 Cergy-Pontoise Cedex, France}

\begin{abstract}
The magnetic phase transition  is experimentally known to give
rise to an anomalous temperature-dependence of the electron
resistivity in ferromagnetic crystals. Phenomenological theories
based on the interaction between itinerant electron spins and
lattice spins have been suggested to explain these observations.
In this paper, we show by extensive Monte Carlo (MC) simulation
the behavior of the resistivity of the spin current calculated as
a function of temperature ($T$) from low-$T$ ordered phase to
high-$T$ paramagnetic phase in a ferromagnetic film.  We analyze
in particular effects of film thickness, surface interactions and
different kinds of impurities on the spin resistivity across the
critical region. The origin of the resistivity peak near the phase
transition is shown to stem from the existence of magnetic domains
in the critical region. We also formulate in this paper a theory
based on the Boltzmann's equation in the relaxation-time
approximation. This equation can be solved using numerical data
obtained by our simulations. We show that our  theory is in a good
agreement with our MC results.  Comparison with experiments is
discussed.

\end{abstract}
 \maketitle
\section{Introduction}
The electric resistivity in magnetic metals has been studied for
years. The main effect of spin-independent resistivity is
unanimously attributed to phonons. As far as spin-dependent
resistivity is concerned, we had to wait until de Gennes and
Friedel's first explanation in 1958\cite{DeGennes} which was based
on the interaction between spins of conduction electrons and
magnetic lattice ions. Experiments have shown that the resistivity
indeed depends on the spin
orientation.\cite{Fert-Campbell,Shwerer,Stishov,Stishov2,Matsukura}
Therefore, the resistivity was expected to depend  strongly on the
spin ordering of the system. Experiments on various magnetic
materials have found in particular an anomalous behavior of the
resistivity at the critical temperature where  the system
undergoes the ferromagnetic-paramagnetic phase
transition.\cite{Shwerer,Stishov,Stishov2,Matsukura}
 The problem of spin-dependent transport has been also extensively
studied in magnetic thin films and multilayers. The so-called
giant magnetoresistance (GMR) was discovered experimentally twenty
years ago.\cite{Baibich,Grunberg} Since then, intensive
investigations, both experimentally and theoretically, have been
carried out.\cite{Fert,review} The so-called "spintronics" was
born with spectacular rapid developments in relation with
industrial applications.  For  recent overviews, the reader is
referred to Refs. \onlinecite{Dietl} and \onlinecite{Barthe}.
Theoretically, in their pioneer work, de Gennes and
Friedel\cite{DeGennes} have suggested that the magnetic
resistivity is proportional to the spin-spin correlation. In other
words, the spin resistivity should behave as the magnetic
susceptibility.  This explained that the resistivity singularity
is due to "long-range" fluctuations of the magnetization observed
in the critical region. Craig  et al\cite{Craig} in 1967 and
Fisher and Langer\cite{Fisher} in 1968 criticized this explanation
and suggested that the shape of the singularity results mainly
from "short-range" interaction at $T\gtrsim T_c$ where $T_c$ is
the transition temperature of the magnetic crystal. Fisher and
Langer have shown in particular that the form of the resistivity
cusp depends on the interaction range. An interesting summary was
published in 1975 by Alexander and coworkers\cite{Alexander} which
highlighted the controversial issue. To see more details on the
magnetic resistivity, we quote an interesting recent publication
from Kataoka.\cite{Kataoka} He calculated the spin-spin
correlation function using the mean-field approximation and  he
could analyze the effects of magnetic-field, density of conduction
electron, the interaction range, etc. Although many theoretical
investigations have been carried out, to date very few Monte Carlo
(MC) simulations have been performed regarding the temperature
dependence of the dynamics of spins participating in the current.
In a recent work,\cite{Akabli} we have investigated by MC
simulations the effects  of magnetic ordering on the spin current
in magnetic multilayers.  Our results are in qualitative agreement
with measurements.\cite{Brucas}

Due to  a large number of parameters which play certainly
important roles at various degrees in the behavior of the spin
resistivity, it is difficult to treat all parameters at the same
time.  The first question is of course  whether the explanation
provided by de Gennes and Friedel can be used in some kinds of
material and that by Fisher and Langer can be applied in some
other kinds of material. In other words, we would like to know the
validity of each of these two arguments.  We will return to this
point in subsection \ref{Discuss}. The second question concerns
the effects of magnetic or non-magnetic impurities on the
resistivity. Note that in 1970, Shwerer and Cuddy\cite{Shwerer}
have shown and compared their experimental results with the
different existing theories to understand the impurity effect on
the magnetic resistivity. However, the interpretation was not
clear enough at the time to understand the real physical mechanism
lying behind. The third question concerns the effects of the
surface on the spin resistivity in thin films. These questions
have motivated the present work.

 In this paper we use extensive MC simulation to study the
transport of itinerant electrons traveling in a ferromagnetic thin
film. We use the Ising model and take into account various
interactions between lattice spins and itinerant spins. We show
that the magnetic resistivity depends on the lattice magnetic
ordering. We analyze this behavior by using a new idea:  instead
of calculating the spin-spin correlation, we calculate the
distribution of clusters in the critical region. We show that the
resistivity depends on the number and the size of clusters of
opposite spins.  We establish also a Boltzmann's equation which
can be solved using numerical data  for the cluster distribution
obtained by our MC simulation.

 The paper is organized as follows. Section II is devoted
to the description of our model and the rules that govern its
dynamics. We take into account (i) interactions between itinerant
and lattice spins, (ii) interactions between itinerant spins
themselves and (iii) interactions between lattice spins. We
include a thermodynamic force due to the gradient of itinerant
electron concentration, an applied electric field and the effect
of a magnetic field. For impurities, we take the
Rudermann-Kittel-Kasuya-Yoshida (RKKY) interaction between them.
In section III, we describe our MC method and discuss the results
we obtained. We develop in section IV a semi-numerical theory
based on the Boltzmann's equation. Using the results obtained with
Hoshen and Kopelman's\cite{Hoshen} cluster-counting algorithm, we
show an excellent agreement between our theory and our MC data.
Concluding remarks are given in Section V.

\section{Interactions and Dynamics}

\subsection{Interactions}
We consider in this paper a ferromagnetic thin film. We use the
Ising model and the face-centered cubic (FCC) lattice with size
 $4N_x\times N_y \times N_z$.  Periodic boundary
conditions (PBC) are used in the $xy$ planes. Spins localized at
FCC lattice sites are called "lattice spins"
 hereafter.  They interact with each other
through the following Hamiltonian

\begin{equation}
\mathcal H_l=-J\sum_{\left<i,j\right>}\mathbf S_i\cdot\mathbf S_j,
\label{eqn:hamil1}
\end{equation}
where $\mathbf S_i$ is the Ising spin at lattice site $i$,
$\sum_{\left<i,j\right>}$ indicates the sum over every
nearest-neighbor (NN) spin pair $(\mathbf S_i, \mathbf S_j)$, $J
(>0)$ being the NN interaction.


In order to study the spin transport in the above system, we
consider a flow of itinerant spins interacting with each other and
with the lattice spins.  The interaction between itinerant spins is
defined  as follows,

\begin{equation}
\mathcal H_m=-\sum_{\left<i,j\right>}K_{i,j}\mathbf
s_i\cdot\mathbf s_j,  \label{eqn:hamil2}
\end{equation}
where $\mathbf s_i$ is the itinerant Ising spin at position  $\vec
r_i$, and $\sum_{\left<i,j\right>}$ denotes a sum over every spin
pair $(\mathbf s_i, \mathbf s_j)$.  The interaction $K_{i,j}$
depends on the distance between the two spins, i.e. $r_{ij}=|\vec
r_i-\vec r_j|$.  A specific form of $K_{i,j}$ will be chosen
below.  The interaction between itinerant spins and lattice spins
is given by

\begin{equation}
\mathcal H_r=-\sum_{\left<i,j\right>}I_{i,j}\mathbf
s_i\cdot\mathbf S_j,  \label{eqn:hamil3}
\end{equation}
where the interaction $I_{i,j}$ depends on the distance
between the itinerant spin $\mathbf s_i$ and the lattice spin
$\mathbf S_i$. For the sake of simplicity, we assume the same form
for $K_{i,j}$ and $I_{i,j}$, namely,
\begin{eqnarray}
K_{i,j}&=& K_0\exp(-r_{ij})\label{eqn:hamil5}\\
I_{i,j}&=& I_0\exp(-r_{ij})\label{eqn:hamil6}
\end{eqnarray}
where $K_0$ and $I_0$ are constants.

\subsection{Dynamics}
\begin{figure}[htp!]
\centering
\includegraphics[width=40mm,height=80mm,angle=-90]{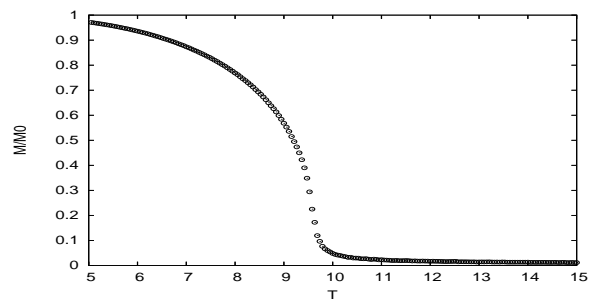}
\caption{Lattice magnetization versus temperature $T$ for $N_z=8$.
$T_c$ is $\simeq 9.58$ in unit of $J=1$.}\label{fig:M(T)}
\end{figure}

The procedure used in our simulation is described as follows.
First we study the thermodynamic properties of the film alone,
i.e. without itinerant spins, using Eq. (\ref {eqn:hamil1}). We
perform MC simulations to determine quantities as the internal
energy, the specific heat, layer magnetizations, the
susceptibility, ... as functions of temperature $T$.\cite{Binder}
From these physical quantities we determine the critical
temperature $T_c$ below which the system is in the ordered phase.
We show in Fig. \ref{fig:M(T)} the lattice magnetization versus
$T$ for $N_z=8$, $N_x=N_y=20$.

Once the lattice has been equilibrated at $T$, we inject $N_0$
itinerant spins into the system. The itinerant spins move into the
system at one end, travel in the $x$ direction, escape the system
at the other end to reenter again at the first end under the PBC.
Note that the PBC are used to ensure that the average density of
itinerant spins remains constant with evolving time (stationary
regime). The dynamics of itinerant spins is governed by the
following interactions:

i) an electric field $\mathbf E$ is applied in the $x$ direction.
Its energy is given by
\begin{equation}
\mathcal {H}_E=-e\mathbf E \cdot \mathbf v,
\end{equation}\
where $ \mathbf v$ is the velocity of the itinerant spin, $e$ its
charge;

ii) a chemical potential term which depends on the concentration
of itinerant spins ("concentration gradient" effect). Its form is
given by
\begin{equation}
\mathcal {H}_{c}= Dn(\mathbf r),
\end{equation}
 where $n(\mathbf r)$ is the concentration of
itinerant spins in a sphere of radius $D_2$ centered at $\mathbf r$.
$D$ is a constant taken equal to $K_0$ for simplicity;

iii) interactions between a given itinerant spin and lattice spins
inside a sphere of radius $D_1$ (Eq.~\ref{eqn:hamil3});

iv) interactions between a given itinerant spin and other
itinerant spins inside a sphere of radius $D_2$
(Eq.~\ref{eqn:hamil2}).

Let us consider the case without an applied magnetic field. The
simulation is carried out as follows: at a given $T$ we calculate
the energy of an itinerant spin by taking into account all the
interactions described above.  Then we tentatively move the spin
under consideration to a new position with a step of length $v_0$ in
an arbitrary direction. Note that this move is immediately rejected
if the new position is inside a sphere of radius $r_0$ centered at a
lattice spin or an itinerant spin. This excluded space emulates the
Pauli exclusion principle in the one hand, and the interaction with
lattice phonons on the other hand.  If the new position does not lie
in a forbidden region of space, then the move is accepted with a
probability given by the standard Metropolis algorithm.\cite{Binder}

To study the case with impurities, we replace randomly a number of
lattice spins $S$  by  impurity spins $\sigma$. The impurities
interact with each other via the RKKY interaction as follows
\begin{equation}
{\mathcal
H}_I=-\sum_{\left<i,j\right>}L(r_i,r_j)\sigma_i\cdot\sigma_j
\label{eqn:hamilI1}
\end{equation}
where
\begin{equation}
L(r_i,r_j)=L_0\cos(2k_F|r_i-r_j|)/|r_i-r_j|^3 \label{eqn:hamilI2}
\end{equation}
 $L_0$ being a constant and $k_F$ the  Fermi wave number of the
lattice.  The impurity spins also interact  with NN lattice spins.
However, to reduce the number of parameters, we take this
interaction equal to $J=1$ as that between NN lattice spins (see
Eq. \ref{eqn:hamil1}) with however $\sigma\neq S$. We will
consider two cases $\sigma=2$ and $\sigma=0$ in this paper.

\section{Monte Carlo Results}

We let $N_0$ itinerant spins travel through the system several
thousands times until a steady state is reached. The parameters we
use in most calculations, except otherwise stated (for example, in
subsection \ref{surf} for $N_z$) are $s=S=1$ and $N_x= N_y=20$ and
$N_z= 8$. Other parameters are $D_1=D_2=1$ (in unit of the FCC
cell length), $K_0=I_0=2$, $L_0=17$, $N_0=8\times 20^2$ (namely
one itinerant spin per FCC unit cell), $v_0=1$,
$k_F=(\dfrac{\pi}{a})(\dfrac{n_0}{2})^{1/3}$, $r_0=0.05$. At each
$T$ the equilibration time for the lattice spins lies around
$10^6$ MC steps per spin and we compute statistical averages over
$10^6$ MC steps per spin. Taking $J=1$, we find that $T_c\simeq
9.58$ for the critical temperature of the lattice spins (see Fig.
\ref{fig:M(T)}).

We define the resistivity $\rho$ as
\begin{equation}
\rho=\frac{1}{n},
\end{equation}
where $n$ is the number of itinerant spins crossing a unit area
perpendicular to the $x$ direction per unit of time.


\subsection{Effect of thickness and effect of magnetic field.}
\begin{figure}[htp!]
\centering
\includegraphics[width=40mm,height=80mm,angle=-90]{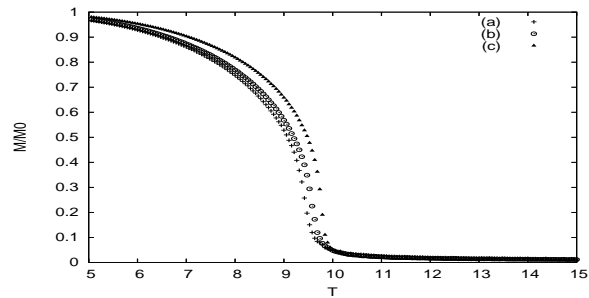}
\caption{Lattice magnetization versus temperature $T$ with
different thicknesses ($N_z$) of the film: crosses, void circles
and black triangles  indicate data for $N_z$=5, 8 and bulk,
respectively. $T_c(bulk)\simeq 9.79$,  $T_c(N_z=8)\simeq 9.58$ and
$T_c(N_z=5)\simeq 9.47$.}\label{M_size}
\end{figure}

\begin{figure}
\includegraphics[width=40mm,height=80mm,angle=-90]{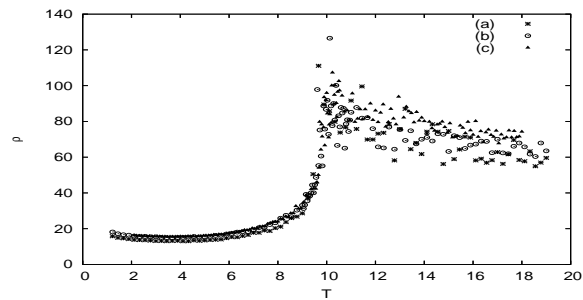}
\caption{Resistivity $\rho$ in arbitrary unit versus temperature
$T$ for different film thicknesses. Crosses (a), void circles (b)
and black triangles (c) indicate data for bulk, $N_z$=8 and 5,
respectively. }\label{R_size}
\end{figure}

 We show in Figs. \ref{M_size} and \ref{R_size} the
simulation results for different thicknesses.  In all cases, the
resistivity $\rho$ is very small at low $T$, undergoes a huge peak
in the ferromagnetic-paramagnetic  transition region, decreases
slowly at high $T$.

We point out that the peak position of the resistivity follows the
variation of critical temperature with changing thickness (see
Fig.~\ref{M_size}) and $\rho$ at $T\gtrsim T_c$ becomes larger
when the thickness decreases.  This is due to the fact that
surface effects tend to slow down itinerant spins.  We return to
this point in the next subsection.

The temperature of resistivity's peak at  a given thickness is
always slightly higher than the corresponding $T_c$.

Let us discuss the temperature dependence of $\rho$ shown in
Fig.~\ref{R_size}:

i) First, $\rho$ is very low in the ordered phase.   We can
explain this by the following argument: below the transition
temperature, there exists a single large cluster of lattice spins
with some isolated "defects" (i. e. clusters of antiparallel
spins), so that any itinerant spin having the parallel orientation
goes through the lattice without hindrance. The resistance is thus
very small but it increases as the number and the size of "defect"
clusters increase with increasing temperature.

ii) Second, $\rho$ exhibits a cusp at the transition temperature.
We present here three interpretations of the existence of this
cusp. Note that these different pictures are not contradictory
with each other. They are just three different manners to express
the same physical mechanism.  The first picture consists in saying
that the cusp is due to the critical fluctuations in the phase
transition region. We know from the theory of critical phenomena
that there is a critical region around the transition temperature
$T_c$. In this region, the mean-field theory should take into
account critical fluctuations. The width of this region is given
by the Ginzburg criterion. The limit of this "Ginzburg" region
could tally with the resistivity's peak and Ginzburg
temperature.\cite{Alexander} The second picture is due to
Fisher-Langer\cite{Fisher} and Kataoka\cite{Kataoka} who suggested
that the form of peak is due mainly to
 short-range spin-spin correlation.  These short-range fluctuations are known to exist
 in the critical region around the critical point.  The third picture comes
 from our MC simulation\cite{Akabli} which showed that the resistivity's peak is due to
 the formation of antiparallel-spin clusters of sizes of a  few lattice cells
 which are known to exist when
 one enters the critical region.
Note in addition that the cluster size is now comparable with the
radius $D_1$ of the interaction sphere, which in turn reduces the
height of potential energy barriers. We have checked this
interpretation by first creating an artificial structure of
alternate clusters of opposite spins and then injecting itinerant
spins into the system.  We observed that itinerant spins do
advance indeed more slowly than in the completely disordered phase
(high-$T$ paramagnetic phase). We have next calculated directly
the cluster-size distribution as a function of $T$ using the
Hoshen-Kopelman's algorithm.\cite{Hoshen} The result confirms the
effect of clusters on the spin conductivity.   The reader is
referred to our previous work\cite{Akabli} for results of a
multilayer case.  We will show in the next section a cluster
distribution for the film studied in this paper.

iii) Third, $\rho$ is large in the paramagnetic phase and
decreases with an increasing temperature. Above $T_c$ in the
paramagnetic phase, the spins become more disordered as $T$
increases: small clusters will be broken into single disordered
spins, so that there is no more energy barrier between successive
positions of itinerant spins on their trajectory. The resistance,
though high, is decreasing with increasing $T$ and saturated as
$T\rightarrow \infty$.

iv) Let us touch upon the effects of varying $D_1$ and $D_2$ at a
low temperatures.  $\rho$ is very small at small $D_1$ ($D_1<0.8$):
this can be explained by the fact that for such small
 $D_1$, itinerant spins do not "see" lattice spins in their interaction sphere so they move almost
 in an empty space.  The effect of $D_2$ is on the other hand qualitatively very different from
 that of $D_1$: $\rho$ is very small at
  small $D_2$ but it increases to very high value  at
 large $D_2$.  We conclude that both $D_1$ and $D_2$ dominate $\rho$
 at their small  values. However, at large values, only $D_2$ has
 a strong effect on $\rho$. This effect comes naturally from the criterion on the itinerant spins concentration
 used in the moving procedure.  Also, we have studied the effect of the electric field $E$ both
above and below $T_c$. The low-field spin current verifies the Ohm
regime. These effects have been also observed in magnetic
multilayer.\cite{Akabli}
 The reader is referred to that work for a detailed presentation of
 these points.

\begin{figure}
\includegraphics[width=40mm,height=80mm,angle=-90]{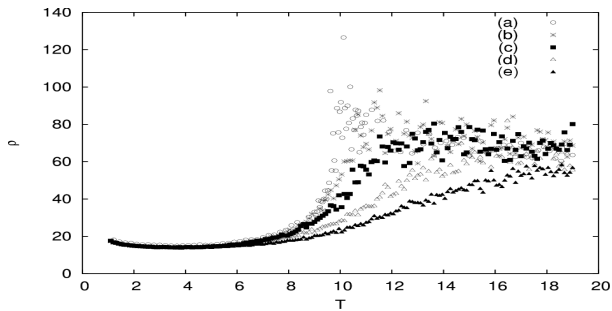}
\caption{Resistivity $\rho$ in arbitrary unit versus temperature
$T$, for different magnetic fields. Void circles, stars, black
rectangles, void triangles and black triangles indicate,
respectively,  data for (a) $B=0$,(b) $B=0.1 J$,(c) $B=0.3 J$,(d)
$B=1 J$ and (e) $B=2 J$.}\label{R_B}
\end{figure}

Let us show now the effect of magnetic field on $\rho$.  As it is
well known, when a magnetic field is applied on a ferromagnet, the
phase transition is suppressed because the magnetization will
never tend to zero.  Critical fluctuations are reduced, the number
of clusters of antiparallel spins diminishes.  As a consequence,
we expect that the peak of the resistivity will be reduced and
disappears at high fields.  This is what we observed in
simulations.  We show results of $\rho$ for several fields in Fig.
\ref{R_B}.

\subsection{Effect of surface}\label{surf}

\begin{figure}
\includegraphics[width=40mm,height=80mm,angle=-90]{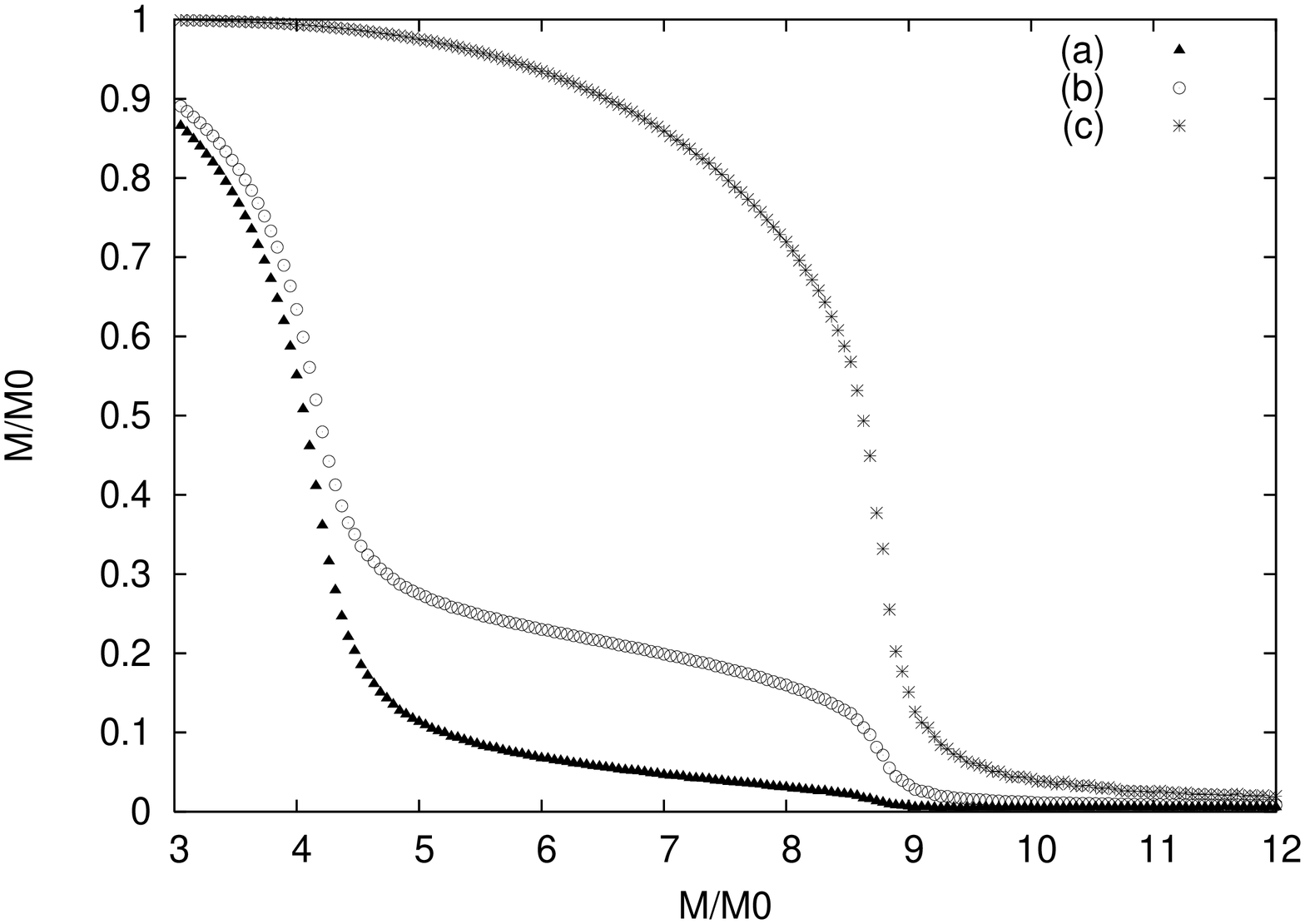}
\includegraphics[width=40mm,height=80mm,angle=-90]{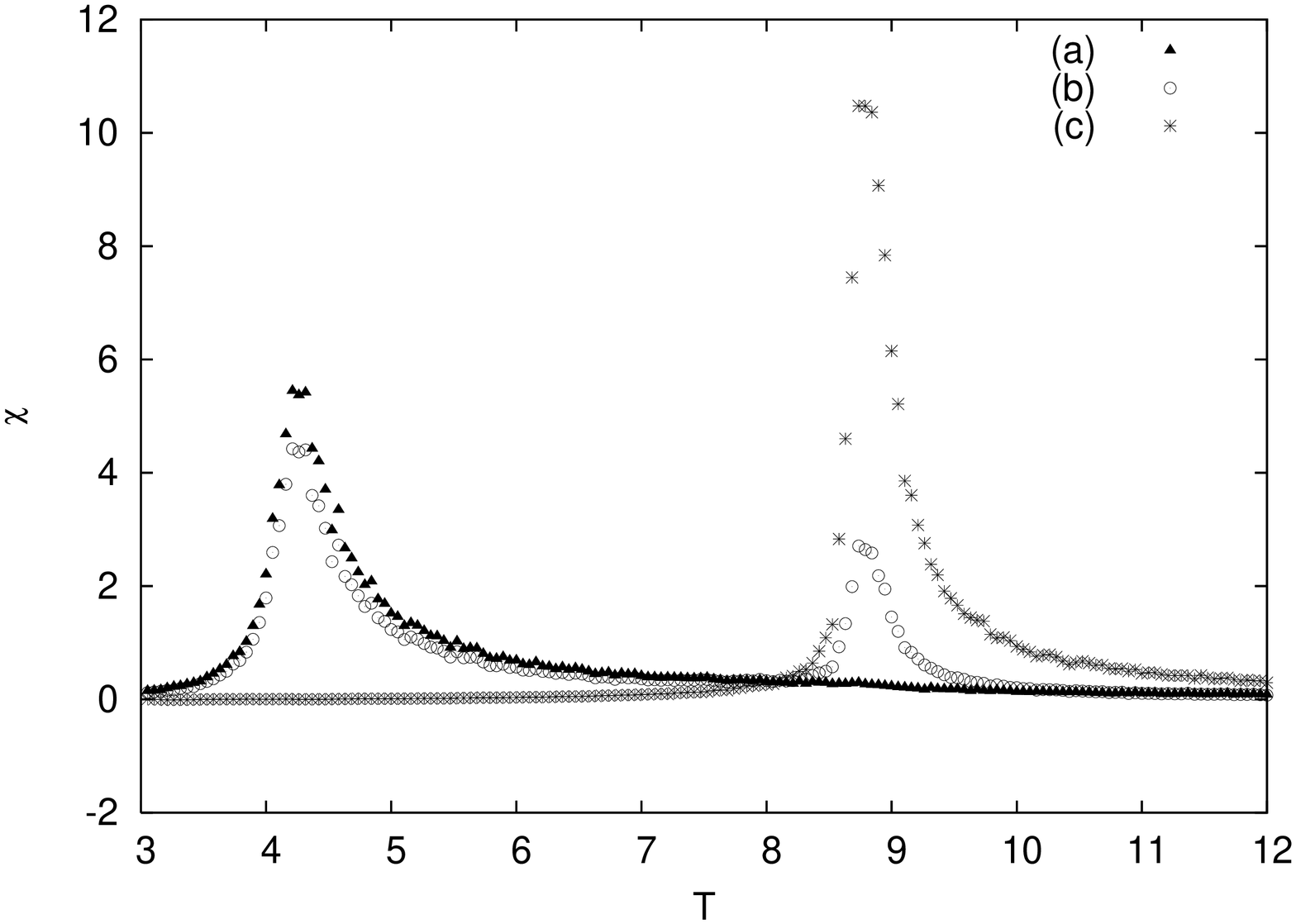}
\caption{Upper figure: Magnetization  versus $T$ in the case where
the system is made of three films: the first and the third have 5
layers with a weaker interaction $J_s$, while the middle has 4
layers with interaction $J=1$. We take $J_s=0.2 J$.  Black
triangles: magnetization of the surface films, stars:
magnetization of the middle film, void circles: total
magnetization. Lower figure: Susceptibility versus $T$ of the same
system as in the upper figure. Black triangles: susceptibility of
the  surface films, stars: susceptibility of the middle films,
void circles: total susceptibility. See text for
comments.}\label{surf1}
\end{figure}

\begin{figure}
\includegraphics[width=40mm,height=80mm,angle=-90]{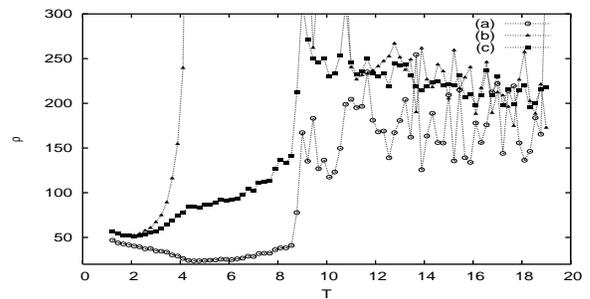}
\caption{Resistivity $\rho$ in arbitrary unit versus  $T$ of the
system described in the previous figure's caption. Black
triangles: resistivity of the surface films, void circles:
resistivity of the middle film,  black squares: total resistivity.
See text for comments.}\label{surf2}
\end{figure}

The picture suggested above on the physical mechanism causing the
variation of the resistivity helps to   understand the surface
effect shown here.  Since the surface spins suffer more
fluctuations due to the lack of neighbors, we expect that surface
lattice spins will scatter more strongly itinerant spins than the
interior lattice spins.  The resistivity therefore should be
larger near the surface.  This is indeed what we observed.  The
effect however is very small in the case where only a single
surface layer is perturbed. To enhance the surface effect, we have
perturbed a number of layers near the surface: we considered a
sandwich of three films: the middle film of 4 layers is placed
between two surface films of 5 layers each.  The in-plane
interaction between spins of the surface films is taken to be
$J_s$ and that of the middle film is $J$. When $J_s=J$ one has one
homogeneous 14-layer film. We have simulated the two cases where
$J_s=J$ and $J_s=0.2J$ for sorting out the surface effect. In the
absence of itinerant spins, the lattice spins undergo a single
phase transition at $T_c\simeq 9.75$ for $J_s=J$, and two
transitions when $J_s=0.2J$: the first transition occurs at
$T_1\simeq 4.20$ for "surface" films and the second at $T_2\simeq
9.60$ for "middle" film. This is seen in Fig. \ref{surf1} where
the magnetization of the surface films drops  at $T_1$ and the
magnetization of the middle film remains up to $T_2$.  The
susceptibility has  two peaks in the case $J_s=0.2J$. The
resistivity of this case is shown in Fig. \ref{surf2}:  at $T<T_1$
the whole system is ordered, $\rho$ is therefore small. When
$T_1\leq T\leq T_2$ the surface spins are disordered while the
middle film is still ordered: itinerant spins encounter strong
scattering in the two surface films, they "escape", after multiple
collisions, to the middle film.   This explains the peak of the
surface resistivity at $T_1$. Note that already far below $ T_1$,
a number of surface itinerant spins begin  to escape to the middle
film, making the resistivity of the middle film to decrease with
increasing $T$ below $T_1$ up to $T_2$, as seen in Fig.
\ref{surf2}.  Note that there is a small shoulder of the total
resistivity at $T_1$. In addition, in the range of temperatures
between $T_1$ and $T_2$ the spins travel almost in the middle film
with a large density resulting in a very low resistivity of the
middle film. For $T>T_2$, itinerant spins flow in every part of
the system.

\subsection{Effect of impurity}

In this subsection, we take back $N_x= N_y=20$ and $N_z= 8$.

\subsubsection{Magnetic impurities}
\begin{figure}[htp!]
\centering
\includegraphics[width=40mm,height=80mm,angle=-90]{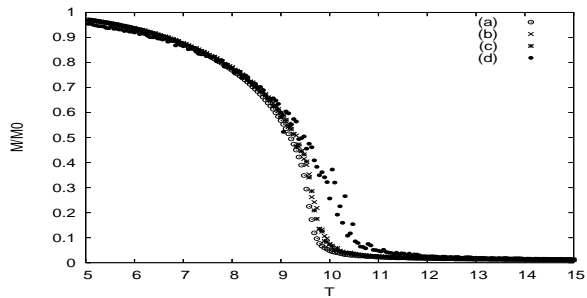}
\caption{Lattice magnetization versus temperature $T$ with
different concentrations of magnetic impurities $C_I$.  Void
circles, crosses, stars and black diamonds indicate, respectively,
data for  (a) $C_I=0\%$, (b) $C_I=1\%$, (c) $C_I=2\%$ and (d)
$C_I=5\%$.}\label{M_mag}
\end{figure}

To treat the case with impurities, we replace randomly a number of
lattice spins $S$  by impurity spins  $\sigma=2$. We suppose  an
RKKY interaction between impurity spins (see Eq.
\ref{eqn:hamilI1}). Figure ~\ref{M_mag} shows the lattice
magnetization for several  impurity concentrations. We see that
critical temperature $T_c$ increases with magnetic impurity's
concentration. We understand that large-spin impurities must
reinforce the magnetic order.

\begin{figure}
\includegraphics[width=40mm,height=80mm,angle=-90]{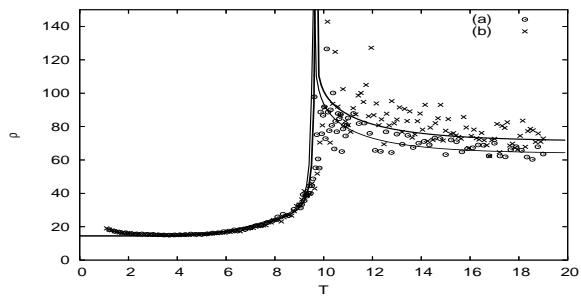}
\caption{Resistivity $\rho$ in arbitrary unit versus temperature
$T$.
 Two cases are shown:  without (a) and with $1\%$ (b)
of magnetic impurities (void circles and crosses, respectively).
Our result using the Boltzmann's equation is shown by the
continuous curves (see sect. IV): thin and thick lines are for (a)
and (b), respectively. Note that $T_c\simeq 9.68$ for
$C_I=1\%$.}\label{mag1}
\end{figure}

\begin{figure}
\includegraphics[width=40mm,height=80mm,angle=-90]{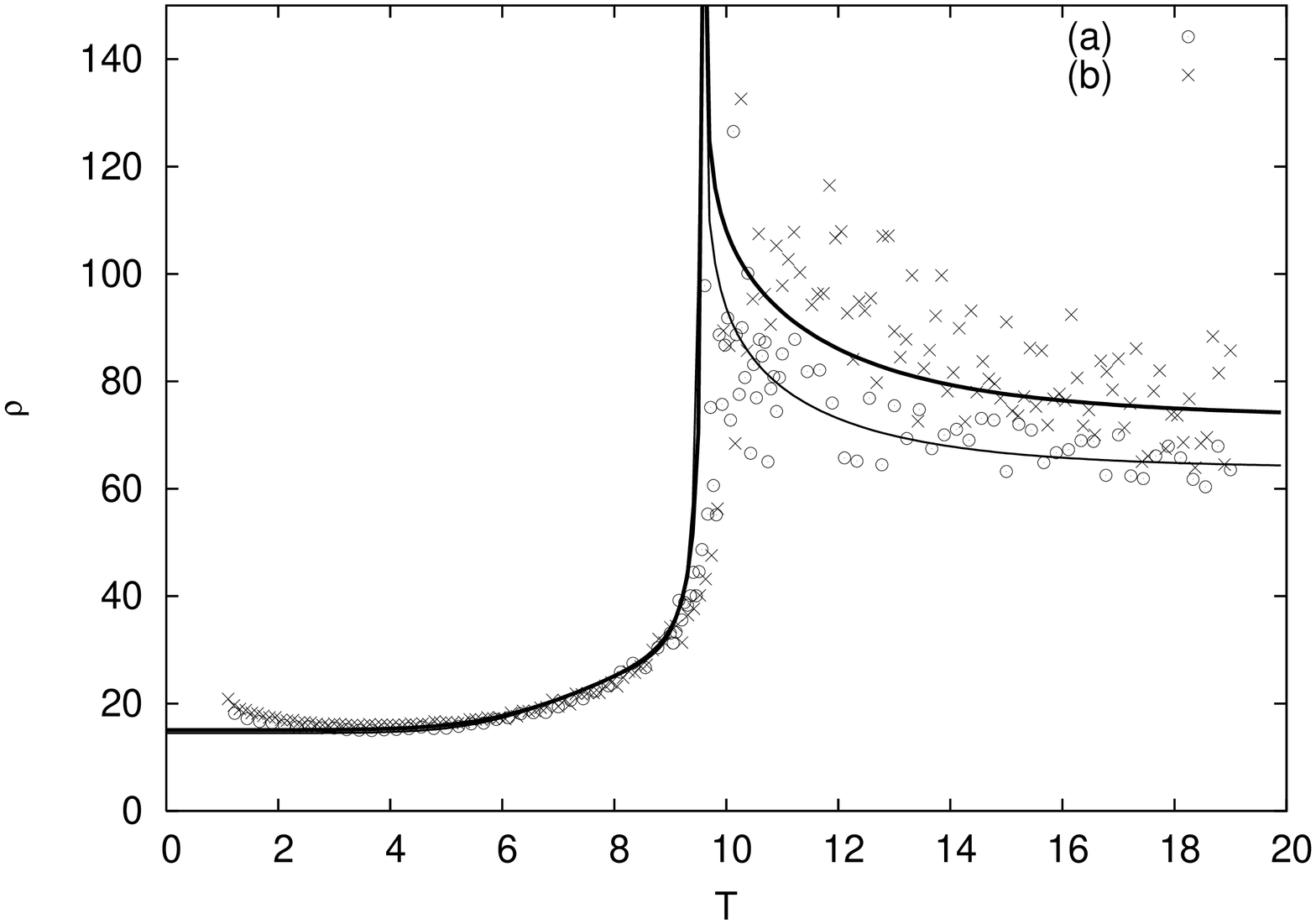}
\caption{Resistivity $\rho$ in arbitrary unit versus temperature
$T$. Two cases are shown:  without  and with $2\%$ of magnetic
impurities (void circles and crosses, respectively). Our result
using the Boltzmann's equation is shown by the continuous curves
(see sect. IV): thin and thick lines are for (a) and (b),
respectively. $T_c\simeq 9.63$ for $2\%$ .}\label{mag2}
\end{figure}

\begin{figure}
\includegraphics[width=40mm,height=80mm,angle=-90]{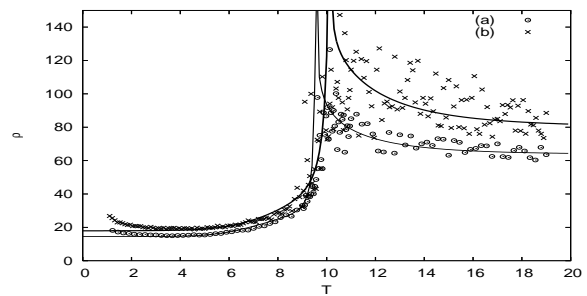}
\caption{Resistivity $\rho$ in arbitrary unit versus temperature
$T$. Two cases are shown:  without (a) and with $5\%$ of magnetic
impurities (b) (void circles and crosses, respectively). Our
result using the Boltzmann's equation is shown by the continuous
curves (see sect. IV): thin and thick lines are for (a) and (b),
respectively. $T_c\simeq 10.21$ for $5\%$.}\label{mag5}
\end{figure}

In Figs.~\ref{mag1},  ~\ref{mag2} and ~\ref{mag5}, we compare a
system without impurity to systems with respectively 1 and 2 and 5
percents of impurities. The temperature of resistivity's peak is a
little higher than the critical temperature and we see that the
peak height increases with increasing impurity concentration (see
Fig. ~\ref{mag5}). This is easily explained by the fact that when
large-spin impurities are introduced into the system, additional
magnetic clusters around these impurities are created in both
ferromagnetic  and paramagnetic phases.  They enhance therefore
$\rho$.

\subsubsection{Non-magnetic impurities}
\begin{figure}[htp!]
\centering
\includegraphics[width=40mm,height=80mm,angle=-90]{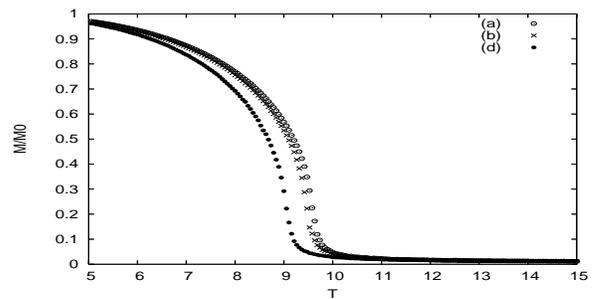}
\caption{Lattice magnetization versus temperature $T$ for
different concentrations of non-magnetic impurities: (a) $C_I=0\%$
(void circles), (b) $C_I=1\%$ (crosses) and (c) $C_I=5\%$ (black
diamonds). }\label{M_nmag}
\end{figure}

\begin{figure}
\includegraphics[width=40mm,height=80mm,angle=-90]{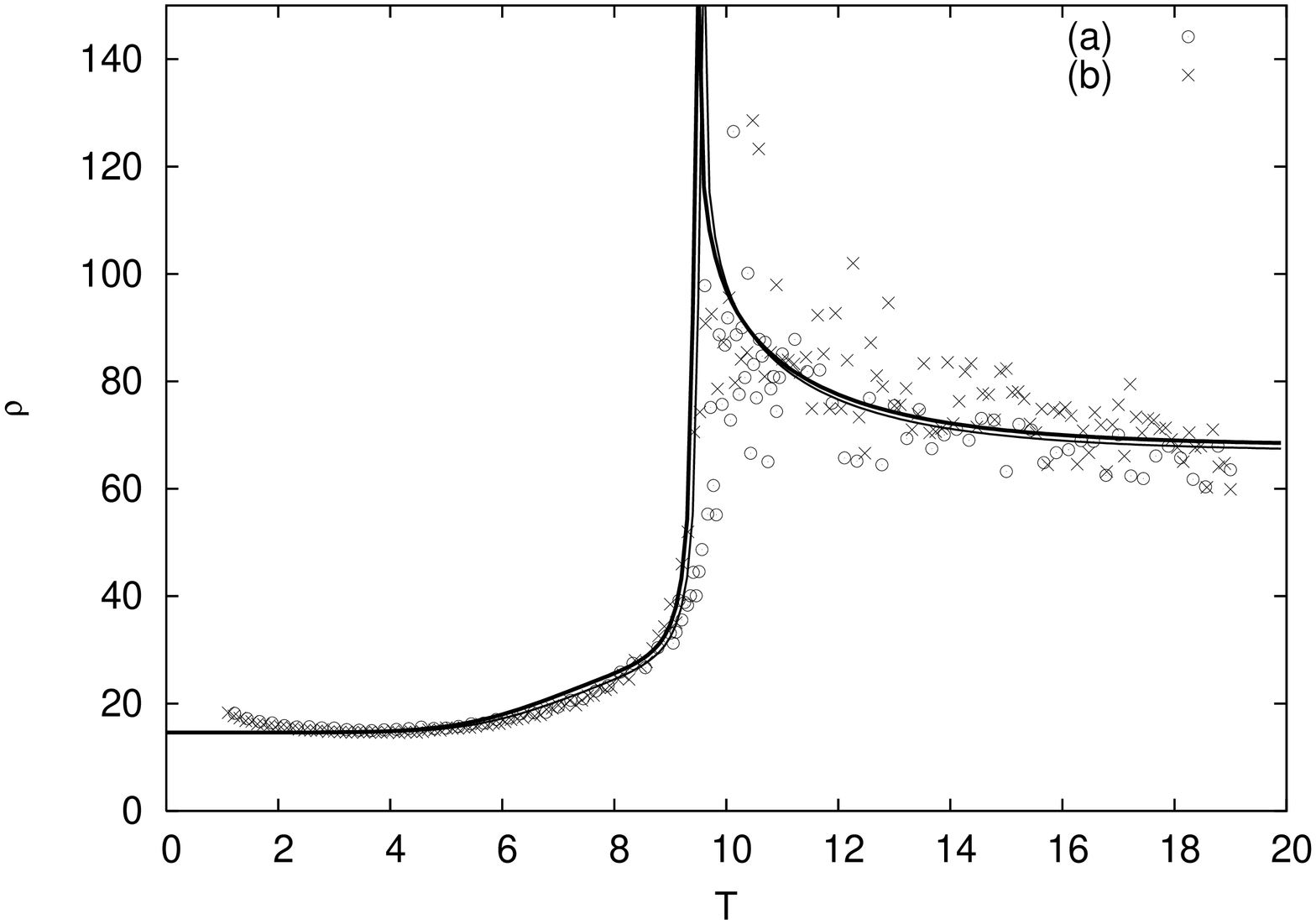}
\caption{Resistivity $\rho$ in arbitrary unit versus temperature
$T$ for two cases: without (a) and with $1\%$ of non-magnetic
impurities (b) (void circles and crosses, respectively). Our
result using the Boltzmann's equation is shown by the continuous
curves (see sect. IV): thin and thick lines are for (a) and (b),
respectively.}\label{nmag1}
\end{figure}

\begin{figure}
\includegraphics[width=40mm,height=80mm,angle=-90]{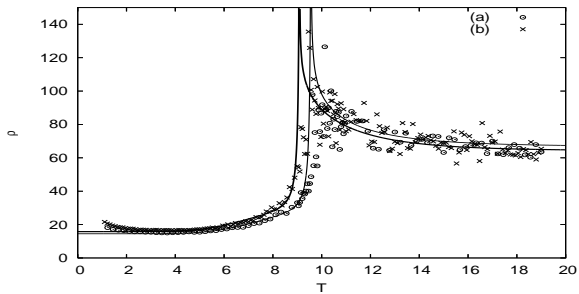}
\caption{Resistivity $\rho$ in arbitrary unit versus temperature
$T$ for two cases: without (a) and with $5\%$ of non-magnetic
impurities (b) (void circles and crosses, respectively). Our
result using the Boltzmann's equation is shown by the continuous
curves (see sect. IV): thin and thick lines are for (a) and (b),
respectively.}\label{nmag5}
\end{figure}

For the case with non-magnetic impurities, we replace randomly a
number of lattice spins $S$ by zero-spin impurities $\sigma=0$.
Figures ~\ref{M_nmag},  \ref{nmag1} and \ref{nmag5} show,
respectively,  the lattice magnetizations  and the resistivities
for non-magnetic impurity concentrations  $1\%$ and $5\%$. We
observe that non-magnetic impurities reduce the critical
temperature and the temperature of the resistivity's peak. This
can be explained by the fact that the now "dilute" lattice spins
has a lower critical temperature so that the scattering of
itinerant spins by lattice-spin clusters should take place at
lower temperatures.

\subsection{Discussion}\label{Discuss}

De Gennes and Friedel\cite{DeGennes} have shown that the
resistivity $\rho$ is related to the spin correlation $<{\bf
S_i}\cdot {\bf S_j}>$.  They have suggested therefore that $\rho$
behaves as the magnetic susceptibility $\chi$. However, unlike the
susceptibility which diverges at the transition, the resistivity
observed in many experiments goes through a finite maximum, i. e.
a cusp, without divergence. To explain this, Fisher and
Langer\cite{Fisher} and then Kataoka\cite{Kataoka} have shown that
the cusp  is due to short-range correlation.  This explanation is
in agreement with many experimental data but not all (see Ref.
\onlinecite{Alexander} for review on early experiments).

Let us recall that $<E>\propto \sum_{i,j}<{\bf S_i}\cdot {\bf
S_j}>$ where the sum is taken over NN (or short-range) spin pairs
while $T\chi \propto <(\sum_i {\bf S_i})^2> =\sum_{i,j}<{\bf
S_i}\cdot {\bf S_j}>$ where the sum is performed over \emph{all}
spin pairs. This is the reason why short-range correlation yields
internal energy and long-range correlation yields susceptibility.

Roughly speaking, if $<{\bf S_i}\cdot {\bf S_j}>$ is short-ranged,
then $\rho$ behaves as $<E>$ so that the temperature derivative of
the resistivity, namely $d\rho/dT$, should behave as the specific
heat with varying $T$. Recent experiments have found this
behavior(see for example Ref. \onlinecite{Stishov2}).

Now, if $<{\bf S_i}\cdot {\bf S_j}>$ is long-ranged, then $\rho$
behaves as the magnetic susceptibility as suggested by de Gennes
and Friedel.\cite{DeGennes} In this case, $\rho$ undergoes a
divergence at $T_c$ as $\chi$. One should have therefore $d\rho
/dT>0$ at $T<T_c$ and $d\rho /dT<0$ at $T>T_c$. In some
experiments, this has been found in for example in magnetic
semiconductors (Ga,Mn)As\cite{Matsukura} (see also Ref.
\onlinecite{Alexander} for review on older experiments). We think
that all systems are not the same because of the difference in
interactions, so one should not discard a priori one of these two
scenarios.

In this paper, we suggest another picture to explain the cusp:
when $T_c$ is approached, large clusters of up (resp. down) spins
are formed in the critical region above $T_c$. As a result, the
resistance is much larger than in the ordered phase: itinerant
electrons have to steer around large clusters of opposite spins in
order to go through the entire lattice. Thermal fluctuations are
not large enough to allow the itinerant spin to overcome the
energy barrier created by the opposite orientation of the clusters
in this temperature region. Of course, far above $T_c$, most
clusters have a small size, the resistivity is still quite large
with respect to the low-$T$ phase. However, $\rho$ decreases as
$T$ is increased because thermal fluctuations are more and more
stronger to help the itinerant spin to overcome energy barriers.

What we have found here is a peak of $\rho$, not a peak of $d\rho
/dT$. So, our resistivity behaves as the susceptibility although
the peak observed here is not sharp and no divergence is observed.
We believe however that, similar to commonly known disordered
systems, the susceptibility peak is broadened more or less because
of the disorder.  The disorder in the system studied here is due
the lack of periodicity in the positions of moving itinerant
spins.

\section{Semi-numerical Theory}

 In this paragraph, we show a theory based on the Boltzmann's equation in the
 relaxation-time approximation.  To solve completely this equation, we shall need some
numerical data from MC simulations for the cluster sizes as will
be seen below. Using these data , we show that our MC result of
resistivity is in a good agreement with this theory.

Let us formulate now the Boltzmann's equation for our system. When
we think about the magnetic resistivity, we think of the
interaction between lattice spins and itinerant spins.  We
recognize immediately the important role of the spin-spin
correlation function in the determination of the mean free-path.
 If we inject through the system a
flow of spins "polarized" in one direction, namely "up", we can
consider clusters of "down" spins in the lattice as "defect
clusters", or as "magnetic impurities", which play the role of
scattering centers. We therefore reduce the problem to the
determination of the number and the size of defect clusters. For
our purpose, we use the Boltzmann's equation with uniform electric
field but without gradient of temperature and gradient of chemical
potential.  We write the equation for $f$, the Fermi-Dirac
distribution function of itinerant electrons, as
\begin{equation}
(\dfrac{\hbar \textbf{k}.e\textbf{E}}{m})(\dfrac{\partial
f^0}{\partial\varepsilon})=(\dfrac{\partial f}{\partial
t})_{coll}, \label{eqn:equ1}
\end{equation}
where $\textbf{k}$ is the wave vector, $e$ and $m$ the electronic
charge and mass, $\epsilon$ the electron energy.  We use the
following relaxation-time approximation
\begin{equation}
(\dfrac{\partial f_k}{\partial t})_{coll}= -
(\dfrac{f_k^{1}}{\tau_k}), \qquad f_k^{1}=f_k-f_k^{0},
\label{eqn:equ2}
\end{equation}
where $\tau_k$ is the relaxation time. Supposing  elastic
collisions, i. e. $k=k'$,  and using the detailed balance we have
\begin{equation}
(\dfrac{\partial f_k}{\partial
t})_{coll}=\dfrac{\Omega}{(2\pi)^{3}}\int
w_{k',k}(f^{1}_{k'}-f^{1}_{k}) dk', \label{eqn:equ3}
\end{equation}
where $\Omega$ is the system volume, $w_{k',k}$ the transition
probability between $k$ and $k'$. We find with
Eq.~(\ref{eqn:equ2}) and Eq.~(\ref{eqn:equ3}) the following
well-known  expression
\begin{eqnarray}
(\dfrac{1}{\tau_k})&=&\dfrac{\Omega}{(2\pi)^{3}}\int
w_{k',k}(1-\cos\theta) \nonumber\\
&&\times \sin\theta k'^{2}dk' d\theta d\phi, \label{eqn:equ4}
\end{eqnarray}
 where $\theta$ and $\phi$ are the angles formed by $\bf{k'}$ with $\bf{k}$, i. e. spherical
 coordinates with $z$ axis parallel to $\bf{k}$.

We use now for Eq.~(\ref{eqn:equ4}) the "Fermi golden rule" and we
obtain
\begin{eqnarray}
(\dfrac{1}{\tau_k})&=&\dfrac{\Omega m}{\hbar^{3}2\pi
k}\int|<k'|V|k>|^{2}(1-\cos\theta)\sin\theta
\nonumber\\
&&\times \delta(k'-k) k'^{2}dk' d\theta \label{eqn:equ5}
\end{eqnarray}
We give for the potential $V$ the following expression which
reminds the form of the interactions
(\ref{eqn:hamil5})-(\ref{eqn:hamil6})
\begin{equation}
 V=V_0\exp(\dfrac{-r}{\xi(T)}), \label{eqn:equ6}
\end{equation}
where $\xi(T)$ is the size of the defect cluster and $V_0$ a
constant. We resolve Eq.~(\ref{eqn:equ5}) with
Eq.~(\ref{eqn:equ6}) and we have the following expression
\begin{equation}
 (\dfrac{1}{\tau_k})=(\dfrac{32V_0^2\Omega mk\pi}{\hbar^{3}})\int
\dfrac{\sin\theta(1-\cos\theta)}{(K^{2}+\xi^{-2})^{3}}d\theta,
\label{eqn:equ7}
\end{equation}
 where $K=|\vec k-\vec k'|$ is given by
\begin{equation}
K=|\vec k-\vec k'|=k[2(1-\cos\theta)]^{1/2}, \label{eqn:equ8}
\end{equation}
Integrating Eq.~(\ref{eqn:equ7}) we obtain
\begin{eqnarray}
(\dfrac{1}{\tau_k})&=&\dfrac{32(V_0\Omega)^2m\pi}{(2k\hbar)^{3}}n_{c}\xi^2\nonumber\\
&&\times
[1-\dfrac{1}{1+4k^2\xi^2}-\dfrac{4k^2\xi^2}{(1+4k^2\xi^2)^2}]
\label{eqn:equ9}
\end{eqnarray}
where $n_c$ is the number of clusters of size $\xi$.   We
integrate Eq.~(\ref{eqn:equ9}) for the interval $[0,k_F]$ and we
obtain finally
\begin{eqnarray}
(\dfrac{1}{\tau})&=&\dfrac{(4V_0\Omega m)^2}{\pi(\hbar)^{3/2}}n_{c}\xi^2\nonumber\\
&&\times[\dfrac{1}{1+(2k_F\xi)^2}+\ln(1+(2k_F\xi)^2)-1]
\label{eqn:equ10}
\end{eqnarray}
Note that although the relaxation time $\tau$ is averaged over all
states in the Fermi sphere, i. e. states at $T=0$,  its
temperature dependence comes from the parameters $\xi$ and $n_c$.
These will be numerically determined in the following. If we know
$\tau$ we can calculate  the resistivity  by the Drude expression
\begin{equation}
\rho_m=\dfrac{m}{ne^{2}}\dfrac{1}{\tau}, \label{eqn:equ11}
\end{equation}

\begin{figure}
\includegraphics[width=40mm,height=80mm,angle=270]{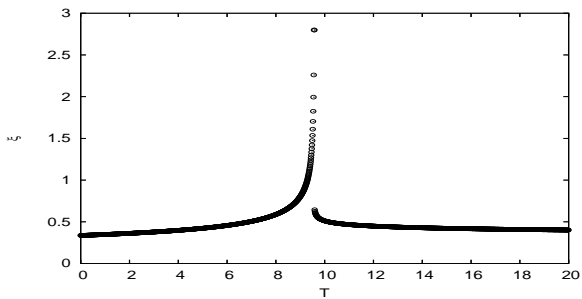}
\caption{Mean size $\xi$ of magnetic clusters versus temperature $T$
for both above and below $T_c$.}\label{rayon}
\end{figure}

\begin{figure}
\includegraphics[width=40mm,height=80mm,angle=270]{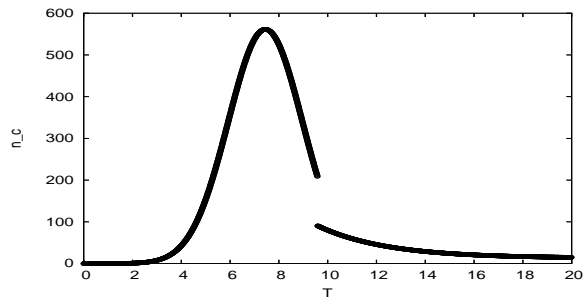}
\caption{Number of average cluster size $n_c$ for both above and
below $T_c$.}\label{nbre_cluster}
\end{figure}

Let us use now the Hoshen-Kopelman's algorithm\cite{Hoshen} to
determine the mean value of $\xi$ and the number of the cluster's
mean size, for different temperatures. The Hoshen-Kopelman's
algorithm allows to regroup into clusters spins with equivalent
value for $T<T_c$ or equivalent energy for $T>T_c$. Using this
algorithm during our MC simulation at a given $T$, we obtain a
"histogram" representing the number of clusters as a function of
the cluster size. For temperature $T$ below $T_c$, we call a
cluster a group of parallel spins surrounded by opposite spins,
and for $T$ above $T_c$ a cluster is a group of spins with the
same energy. At a given $T$, we estimate the average size $\xi$
using the histogram as follows: calling $N_i$ the number of spins
in the cluster and $P_i$ the probability of the cluster deduced
from the histogram, we have
\begin{equation}
\xi=\dfrac{\sum_i N_iP_i}{\sum_i P_i}, \label{eqn:equ12}
\end{equation}
In doing this we obtain  $\xi$ for the whole temperature range. We
note that we can fit the cluster size $\xi$ with the following
formula
\begin{equation}
\xi=A|T_c-T|^{\nu/3}, \label{eqn:equ13}
\end{equation}
where $\nu$ is a fitting parameter and $A$ a constant.  These
parameters are different for $T<T_c$ and $T>T_c$.  Figure
~\ref{rayon} and Figure ~\ref{nbre_cluster} show the average size
and the average number of cluster versus temperature. To simplify
our approach we consider that the cluster's geometry is a sphere
with radius $\xi$.  Note that due to the fact that our fitting was
made separately for $T<T_c$ and $T>T_c$, no effort has been made for
the matching at $T=T_c$ exactly, but this does not affect the
behavior discussed below.

We distinguish hereafter temperatures below and above $T_c$ in
establishing our theory. We write

\[T<T_c
   \left [
   \begin{array}{c}
\rho_m=\rho_0 (1+C_{inf}n_{c}\xi^2[-1+\dfrac{1}{1+(2k_F\xi)^2}\nonumber\\
+\ln(1+(2k_F\xi)^2)]),\\
\xi=(\dfrac{3A_{inf}}{16\pi})^{1/3}(T_c-T)^{\nu_{inf}/3},\\
n_{c}=(\dfrac{B_{inf}}{2\alpha \pi})\times
\exp[\dfrac{-(T-T_G)^2}{2\alpha^2}].
   \end{array}
   \right .
\]
\begin{equation}\label{eqn:equ14}\end{equation}



\[T>T_c
   \left [
   \begin{array}{c}
\rho_m=\rho_\infty (1+C_{sup}n_{c}\xi^2[-1+\dfrac{1}{1+(2k_F\xi)^2}\nonumber\\
+\ln(1+(2k_F\xi)^2)]),\\
\xi=(\dfrac{3A_{sup}}{16\pi})^{1/3}(T-T_c)^{\nu_{sup}/3},\\
n_{c}=B_{sup}\exp[-D(T-T_c)]+n_0.
   \end{array}
   \right .
\]
\begin{equation}\label{eqn:equ17}\end{equation}



In Eq. (\ref{eqn:equ14}) we call $T_G$ the temperature on which
the cluster of small size are gathering to form bigger cluster.
This temperature marks the limit when one  enters the critical
region from below. In Eq. (\ref{eqn:equ14}), $\alpha$ is the
half-width of the peak of $n_c$ shown in Fig.~\ref{nbre_cluster}.
$\rho_0$ is the resistivity at $T=0$ and $\rho_{\infty}$ is that
at $T=\infty$.

We summarize in  Table 1 the different results obtained for the
cases studied by MC simulations shown above. Other parameters
$A_{inf}$, $B_{inf}$, $C_{inf}$, $A_{sup}$, $B_{sup}$ and
$C_{sup}$ are fitting parameters which are not of physical
importance and therefore not given here for the sake of clarity.
Using the numerical values of Table 1 and the average cluster size
and cluster number shown in Fig. ~\ref{rayon} and Fig.
~\ref{nbre_cluster}, we plot
 Eqs. (\ref{eqn:equ14}) and (\ref{eqn:equ17}) by continuous lines in Figs. ~\ref{mag1}-\ref{nmag5}
to compare with MC simulations shown in these figures.
 We emphasize  that our theory provides a good
"fit" for simulation results.

\begin{widetext}
\begin{table*}
\centering
  \caption{Various  numerical values obtained by MC simulations which
are used to plot Eqs. (\ref{eqn:equ14}) and
(\ref{eqn:equ17}).}\label{tab:criexp}
\begin{tabular}{| r | c | c | c | c | c | c | c |}
\hline $\qquad Impurity \qquad $ & $\qquad T_c \qquad$ & $\qquad T_G \qquad$ & $\qquad \nu_{inf}\qquad $ & $\qquad  \nu_{sup}\qquad $ & $\qquad  \alpha\qquad $\\
\hline
$0\%$\qquad S=1 & 9.58 & 7.4443 +/- 0.066& -0.9254 +/- 0.015& -0.2267 +/- 0.006& 1.51875 +/- 0.07\\
$1\%$\qquad $\sigma$=2 & 9.68 & 7.5006 +/- 0.054& -0.9253 +/- 0.017& -0.1449 +/- 0.006& 1.62908 +/- 0.06\\
$2\%$\qquad $\sigma$=2 & 9.63 & 7.7103 +/- 0.049& -0.9856 +/- 0.016& -0.1135 +/- 0.004& 1.64786 +/- 0.05\\
$5\%$\qquad $\sigma$=2 & 10.2 & 7.9658 +/- 0.094& -1.1069 +/- 0.016& -0.0747 +/- 0.002& 2.13618 +/- 0.10\\
$1\%$\qquad $\sigma$=0 & 9.47 & 7.2866 +/- 0.062& -0.9028 +/- 0.013& -0.2106 +/- 0.006& 1.51766 +/- 0.06\\
$5\%$\qquad $\sigma$=0 & 9.10 & 7.0105 +/- 0.054& -0.9381 +/- 0.015& -0.1607 +/- 0.006& 1.47261 +/- 0.05\\
\hline
\end{tabular}
\end{table*}
\end{widetext}

\begin{figure}
\includegraphics[width=40mm,height=80mm,angle=-90]{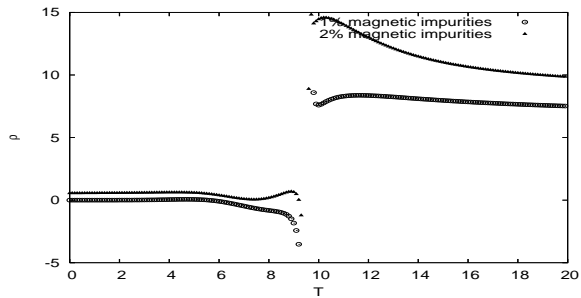}
\caption{Resistivity $\rho_I$ in arbitrary unit versus temperature
$T$. Void circles and black triangles indicate data for $1\%$ and
$2\%$ magnetic impurities, respectively.}\label{romag}
\end{figure}

\begin{figure}
\includegraphics[width=40mm,height=80mm,angle=-90]{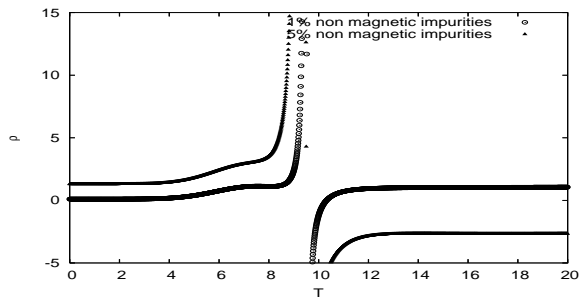}
\caption{Resistivity $\rho_I$ in arbitrary unit versus temperature
$T$. Void circles and black triangles indicate data for $1\%$ and
$5\%$ non magnetic impurities, respectively.}\label{ronmag}
\end{figure}

Based on those results, we can extract the resistivity $\rho_I$
corresponding only to the addition of impurities. $\rho_I$ is
defined as
\begin{equation}
\mathcal \rho_I=\rho_m-\rho_{standard}, \label{eqn:equ20}
\end{equation}
where $\rho_{standard}$ is the resistivity without impurities (see
the first line of Table 1).
  We compare now the
resistivity $\rho_I$ with experiments realized by Shwerer and
Cuddy.\cite{Shwerer} It is important to note that the change of
behavior can be explained if we use the correct value for $\nu$,
$T_c$, $T_G$, etc. Figure ~\ref{romag} shows $\rho_I$ with
magnetic impurities corresponding to the Ni-Fe system , while Fig.
~\ref{ronmag} shows $\rho_I$ in the case of non-magnetic
impurities corresponding to the Ni-Cr case.



We see that the form of $\rho_I$ of the alloys Ni-Fe($1\%$) and
Ni-Fe($0.5\%$) experimentally observed\cite{Shwerer} can be
compared to the curves of $1\%$ and $2\%$ of magnetic impurities
shown in Fig. \ref{romag}. For Ni-Cr($1\%$) and Ni-Cr($2\%$),
experimental curves are in agreement with our results of
non-magnetic impurities shown in Fig. \ref{ronmag}.

\begin{figure}
\includegraphics[width=40mm,height=80mm,angle=-90]{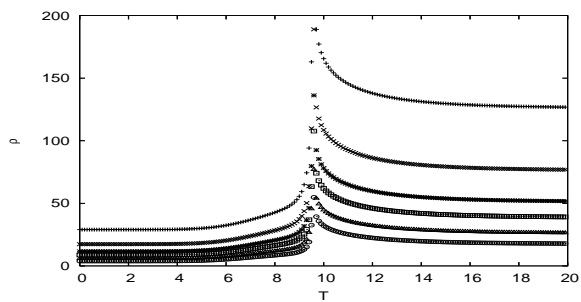}
\caption{Resistivity $\rho$ in arbitrary unit versus temperature
$T$ for different densities of itinerants spins. Curves from top
to bottom are for $n_0=0.2$, 0.33, 0.5, 0.7, 1 and
1.5.}\label{densite}
\end{figure}

Finally, to close this section, let us show theoretically from the
equations obtained above, the effects of the density of itinerant
spins on the resistivity.  Figure \ref{densite} shows that, as the
density $n_0$ is increased, the peak of $\rho$ diminishes.  It is
noted that this behavior is very similar with that obtained by
Kataoka.\cite{Kataoka} In our MC simulation shown above, we have
chosen $n_0=1/4$. Such a weak density has allowed us to avoid the
flip of lattice spins upon interaction with itinerant spins.  In
the case of strong density, we expect that a number of lattice
spins, when surrounded by a large number of itinerant spins,
should flip to accommodate themselves with their moving neighbors.
So the lattice ordering should be affected. As a consequence,
critical fluctuations of lattice spins are more or less
suppressed, so is the peak's height, just like in the case of an
applied magnetic field. Kataoka\cite{Kataoka} has found this in
his calculation  by taking into account the spin flipping: the
resistivity's peak disappears then at the transition. It would be
interesting to perform more MC simulations with varying $n_0$.
This is left for a future investigation.

\section{Conclusion}
We have shown in this paper results of MC simulations on the
transport of itinerant spins interacting with localized lattice
spins in a ferromagnetic FCC thin film. Various interactions have
been taken into account. We found that the spin current is
strongly dependent on the lattice spin ordering: at low $T$
itinerant spins whose direction is parallel  to the lattice spins
yield a strong current, namely a small resistivity. At the
ferromagnetic transition, the resistivity undergoes a huge peak.
At higher temperatures, the lattice spins are disordered, the
resistivity is  still large but it decreases with increasing $T$.
From the discussion given in subsection \ref{Discuss}, we conclude
that the resistivity $\rho$ of the model studied here behaves as
the magnetic susceptibility with a peak at $T_c$.   $d\rho /dT$,
differential resistivity is thus negative for $T>T_c$. The peak of
the resistivity obtained here is in agreement with experiments on
magnetic semiconductors (Ga,Mn)As for example.\cite{Matsukura} Of
course, to compare the peak's shape experimentally obtained for
each material, we need to refine our model parameters for each of
them. This was not the purpose of the present paper. Instead, we
were looking for generic effects to show physical mechanisms lying
behind the temperature dependence of the spin resistivity.  In
this spirit, we note that early theories have related the origin
of the peak to the spin-spin correlation, while our interpretation
here is based on the existence of defect clusters formed in the
critical region. This interpretation has been verified by
calculating the number and the size of clusters as a function of
$T$ by the use of Hoshen-Kopelman's algorithm. We have formulated
a theory based on the Boltzmann's equation.  We solved this
equation using numerical data obtained for the  number and the
size of average cluster at each $T$. The results on the
resistivity are in a good agreement with  MC results.

Finally, let us conclude by saying that the clear physical picture
we provide in this paper for the understanding of the behavior of
the resistivity in a single ferromagnetic film  will help to
understand properties of resistivity in more complicated systems
such as antiferromagnets, non-Ising spin systems, frustrated spin
systems, disordered media, ... where much has to be done.

{}

\end{document}